\newif\ifanon
\anonfalse
\ifanon
\documentclass[sigconf,review,anonymous]{acmart}
\else
\documentclass[sigconf,nonacm]{acmart}
\fi
\ifanon
\setcopyright{none}
\renewcommand\footnotetextcopyrightpermission[1]{}
\acmConference[SenSys '27]{ACM/IEEE International Conference on Embedded
    Artificial Intelligence and Sensing Systems}{May 17--20, 2027}{Boulder,
    CO, USA}
\fi

\ifanon
\newcommand{\simulatorlink}{GitHub repository omitted to preserve author anonymity.}
\else
\newcommand{\simulatorlink}{\url{https://gitlab.ilabt.imec.be/iwine/ra-pdoa-simulator}}
\fi

\newcommand{\compacttable}{%
\addtolength{\tabcolsep}{-0.1em}%
\renewcommand{\arraystretch}{1.1}}

\usepackage{amsmath}
\usepackage{graphicx}
\usepackage{booktabs,makecell}
\usepackage{acronym}
\usepackage{tikz}
\usetikzlibrary{positioning,arrows.meta,fit,backgrounds,calc}

\begin{document}

\title[Ranging-Assisted PDoA for BLE Channel Sounding]{Scalable BLE Channel Sounding Localization with Ranging-Assisted PDoA}

\author{Jorg Wieme}
\orcid{0000-0003-4542-0490}
\email{jorg.wieme@ugent.be}
\affiliation{%
\institution{IDLab, Ghent University -- imec}
\city{Ghent}
\country{Belgium}}

\author{Eli De Poorter}
\orcid{0000-0002-0214-5751}
\email{eli.depoorter@ugent.be}
\affiliation{%
\institution{IDLab, Ghent University -- imec}
\city{Ghent}
\country{Belgium}}

\author{Jeroen Hoebeke}
\orcid{0000-0003-2039-007X}
\email{jeroen.hoebeke@ugent.be}
\affiliation{%
\institution{IDLab, Ghent University -- imec}
\city{Ghent}
\country{Belgium}}

% Header author list for the arXiv build only; the review build keeps
% acmart's "Anon." so no name can reach the running header.
\ifanon
\renewcommand{\shortauthors}{Anon}
\else
\renewcommand{\shortauthors}{Jorg Wieme et al.}
\fi

\begin{abstract}
Bluetooth Channel Sounding (CS) enables accurate phase-based ranging, but multi-anchor localization requires sequential ranging exchanges that limit scalability and increase measurement latency. This paper proposes Ranging-Assisted Phase Difference of Arrival (RA-PDoA), a scalable BLE localization architecture combining one active anchor performing a standard two-way CS exchange with multiple passive receive-only anchors observing the same transmissions. A channel-recovery framework is derived that enables passive anchors to reconstruct a channel response suitable for super-resolution delay estimation, allowing multipath-robust ranging without participating in a two-way exchange. Coarse arrival-time information is further used to bound the position estimate and reject inconsistent range solutions. A comprehensive evaluation characterizes the impact of multipath propagation, body shadowing, synchronization requirements, oscillator impairments, anchor placement, and deployment inaccuracies, thereby quantifying the practical infrastructure requirements for scalable BLE localization. Simulation results in modeled indoor environments show that RA-PDoA reduces the 90th-percentile localization error from 175.3\,cm to 95.5\,cm under body shadowing while maintaining an airtime requirement independent of the number of anchors. The results demonstrate that a single active ranging link enables scalable and multipath-robust BLE Channel Sounding localization.
\end{abstract}

\ifanon

\else
\fi


\acrodef{GPS}{Global Positioning System}
\acrodef{NLOS}{Non-Line-of-Sight}
\acrodef{LOS}{Line-of-Sight}
\acrodef{RF}{Radio Frequency}
\acrodef{UWB}{Ultra-Wideband}
\acrodef{FMCW}{Frequency-Modulated Continuous-Wave}
\acrodef{RSSI}{Received Signal Strength Indicator}
\acrodef{DF}{Direction Finding}
\acrodef{BLE}{Bluetooth Low Energy}
\acrodef{CS}{Channel Sounding}
\acrodef{ISAC}{Integrated Sensing and Communications}
\acrodef{PDoA}{Phase Difference of Arrival}
\acrodef{RA-PDoA-CS}{Ranging-Assisted Phase Difference of Arrival Channel Sounding}
\acrodef{RA-PDoA}{Ranging-Assisted Phase Difference of Arrival}
\acrodef{ToA}{Time of Arrival}
\acrodef{MUSIC}{Multiple Signal Classification}
\acrodef{MVDR}{Minimum Variance Distortionless Response}
\acrodef{MCPD}{Multi-Carrier Phase Difference}
\acrodef{PAwR}{Periodic Advertising with Responses}
\acrodef{TDoA}{Time Difference of Arrival}
\acrodef{DS-TWR}{Double-Sided Two-Way Ranging}
\acrodef{RTT}{Round-Trip Timing}
\acrodef{PBR}{Phase-Based Ranging}
\acrodef{LE}{Low Energy}
\acrodef{PCT}{Phase Correction Term}
\acrodef{LO}{Local Oscillator}
\acrodef{MDL}{Minimum Description Length}
\acrodef{AIC}{Akaike Information Criterion}
\acrodef{SNR}{Signal-to-Noise Ratio}
\acrodef{CEP90}{$90$th-percentile Circular Error Probable}

\maketitle

\acmSubmissionID{657}

\section{Introduction}
Accurate indoor localization supports healthcare, robotics, and logistics
\cite{logistics8040126, app15031544, robotics12020047}. Among the available
radio technologies, \ac{BLE} is inexpensive and widely integrated into
commercial devices, while \ac{UWB} offers higher accuracy but requires dedicated
radio hardware \cite{8692423}. Existing \ac{BLE} approaches make different
tradeoffs: \ac{RSSI} is sensitive to multipath \cite{7103024, smartcities8050153, 9878154, s23208598}, and
\ac{DF} requires antenna arrays \cite{s21113589, 9190573}. \ac{BLE} \ac{CS}
instead estimates range from phase across frequency channels, with round-trip
timing providing a coarse check \cite{ble6}. Commercial evaluations have demonstrated its sub-meter potential \cite{11207662}.

Accurate point-to-point ranges do not, by themselves, provide scalable
multi-anchor localization. Standard \ac{CS} uses a connection-oriented
exchange, so anchors are ranged sequentially, airtime and latency therefore
increase with anchor count \cite{543845}. Indoor multipath creates a separate accuracy problem: within \ac{BLE}'s limited bandwidth, direct and reflected paths combine and bias phase-slope estimates. A passive broadcast could let many anchors observe one transmission, but a receive-only anchor can cancel the unknown tag phase only against other anchors, leaving a ratio of channel responses rather than the channel response required by conventional super-resolution estimators.

We propose \ac{RA-PDoA} to address both limitations. One active anchor performs a standard two-way \ac{CS} exchange, while synchronized passive anchors observe the same transmissions. The active response provides the reference needed to recover a channel response at each passive anchor and estimate its propagation delays with super-resolution methods. Coarse arrival-time information also bounds the position and rejects inconsistent ranges. The ranging airtime per cycle is therefore independent of anchor count. We discuss implementation requirements in Section~\ref{sec:implementation}.

We evaluate \ac{RA-PDoA} in simulation as a feasibility and infrastructure-requirements
study. Under body shadowing, it reduces the 90th-percentile localization error
from $175.3$ to $95.5$~cm compared to broadcast \ac{PDoA} \cite{10.1145/3842657.3842787}. The contributions are:
\begin{itemize}
    \item We introduce \ac{RA-PDoA}, an architecture that uses one two-way \ac{CS} exchange per localization cycle, independent of the number of passive anchors.
    \item We derive a channel-recovery identity that enables super-resolution estimation at passive anchors without additional ranging exchanges.
    \item We combine model-order selection, per-link estimator selection, and coarse arrival-time validation to mitigate multipath and reject inconsistent ranges.
    \item We separate architectural and estimator effects, quantify body-shadowing and multipath performance, and derive synchronization and deployment requirements.
    \item We have made the simulator source code publicly available to support reproduction of the reported results.\footnote{\simulatorlink}
\end{itemize}

The paper reviews related work, establishes the signal model, presents \ac{RA-PDoA} and its estimator, then evaluates the design and its infrastructure requirements.

\section{Related Work} \label{sec:related_work}

\subsection{Scalable Multi-Anchor Ranging}

Multi-anchor localization commonly relies on point-to-point ranging followed by trilateration. In \ac{UWB} systems, \ac{DS-TWR} and its variants provide accurate, self-referenced range measurements \cite{7822844}. However, each anchor requires a separate two-way exchange, causing the required airtime and measurement latency to increase with the number of anchors \cite{10423865, 9600859}. This sequential acquisition is particularly problematic for moving tags, as the ranges used for a single position estimate are obtained at different time instants. 

Passive reception provides a more scalable alternative. In \ac{TDoA} localization, multiple anchors observe a single tag transmission \cite{s23125710}, making the required airtime independent of the number of anchors. This scalability comes at the cost of tightly synchronized anchor clocks, since timing offsets are indistinguishable from propagation-time differences. Active--passive ranging protocols provide an intermediate solution by combining two-way ranging with observations from passive anchors \cite{9600859}. The proposed \ac{RA-PDoA} architecture follows the active--passive principle but applies it to phase-based \ac{BLE} \ac{CS}: one anchor performs a two-way exchange, while the remaining anchors passively observe the same transmissions.

\subsection{Phase-Based Ranging with BLE} Before the standardization of \ac{BLE} \ac{CS}, multicarrier phase-based ranging techniques such as \ac{MCPD} estimated distance from the phase progression across frequency channels. Full-spectrum channel hopping was used to improve robustness against frequency-selective fading and multipath \cite{8885791}. Nevertheless, conventional phase-slope estimation assumes that the measured phase progression corresponds to a single propagation path. Super-resolution estimators have therefore been investigated to recover individual propagation delays from the measured frequency response. \ac{MUSIC}-based processing can improve ranging performance in multipath environments \cite{9128721}, while subsequent adaptations increase robustness when parts of the frequency spectrum are missing or affected by interference \cite{10092848}. However, these methods assume that the channel frequency response between tag and anchor is measured directly, which requires a two-way exchange between both endpoints to cancel the unknown transmitter phase. Following the introduction of standardized \ac{BLE} \ac{CS}, recent studies have focused on estimator design and experimental performance. Processing based on \ac{MVDR} has demonstrated ranging errors below 1.2~m in complex indoor environments \cite{10669801}. Evaluations on commercial hardware further show that \ac{MUSIC}-based processing can achieve sub-meter ranging errors under indoor \ac{LOS} conditions, but also reveal substantial degradation under \ac{NLOS} propagation and unfavorable device orientations \cite{11207662}. Similar limitations have been reported in pedestrian-localization \cite{varughese2026assessinglocalizationtechnologiespedestrian} and keyless-entry applications \cite{10965631}. Online adaptation of the ranging-error distribution has also been proposed to support real-time localization without offline calibration \cite{11275429}. Importantly, existing super-resolution methods cannot be transferred directly to passive phase-difference localization. Cancelling the unknown transmitter phase across synchronized anchors yields a ratio of channel responses rather than the frequency response of an individual link. In this work, \ac{PDoA} therefore refers to phase differences measured across spatially distributed anchors, rather than conventional inter-antenna phase differences within a multi-antenna receiver.

\subsection{Connectionless BLE CS} Connectionless operation has recently been proposed to reduce the connection and scheduling overhead of multi-device \ac{BLE} \ac{CS}. The approach in \cite{schex2026connectionlessbluetoothchannelsounding} uses \ac{PAwR} to coordinate scalable and energy-efficient ranging. It reduces the sequential overhead of conventional connected operation, but relies on \ac{BLE} test-mode commands and does not address the multipath bias of phase-slope ranging. Broadcast \ac{PDoA}, where synchronized passive anchors observe a single tag transmission, was outlined for smart-hospital infrastructure in \cite{10.1145/3842657.3842787}, without a two-way reference link. To summarize, prior work addresses either scalable multi-anchor acquisition or multipath-robust point-to-point ranging.

\section{System Model} \label{sec:system_model}

Throughout, the deployment consists of a tag at an unknown planar coordinate $\mathbf p = (x,y)$ and $N$ synchronized anchors at known positions $\mathbf q_i = (x_i,y_i)$. Anchors are mounted above the tag, so that the propagation distance is three-dimensional while the unknown remains planar,
\begin{equation}
    d_i = \sqrt{\left\| \mathbf p - \mathbf q_i \right\|^2 + \Delta z_i^2},
    \label{eq:geometry}
\end{equation}
where $\Delta z_i$ denotes the height of anchor $i$ above the tag and is taken as known. This reduces the position to two unknowns while retaining the true propagation distance, and applies equally to the two-way procedure of Section~\ref{sec:ble_cs} and the broadcast arrangement of Section~\ref{sec:broadcast_cs}.

\subsection{BLE Channel Sounding} \label{sec:ble_cs}

Introduced in the Bluetooth Core Specification 6.0 \cite{ble6}, \ac{CS} is a two-way ranging framework that operates between an initiator and reflector over an established, encrypted \ac{LE} connection. To achieve reliable distance estimation, \ac{CS} integrates two complementary methods. \ac{PBR} transmits unmodulated continuous-wave tones across a set of channels and recovers the distance from the phase slope over frequency, using a channel spacing of $\Delta f = 1$ MHz. Conversely, \ac{RTT} estimates the range based on the two-way propagation delay of exchanged messages, calculated as $\hat d = c(T_{\mathrm{round}}-T_{\mathrm{reply}})/2$. \ac{RTT} yields a coarse but unambiguous distance estimate, so it primarily serves as a secure bounding mechanism to validate \ac{PBR}.

During \ac{PBR}, the initiator transmits a tone at frequency $f_k$. The reflector returns this tone phase-aligned with the incoming signal, enabling the initiator to measure the received phase coming from the reflector. Combining these one-way measurements cancels the unknown local-oscillator phase offsets, yielding the round-trip phase
\begin{equation}
    \theta_k = \frac{4\pi f_k d}{c} + 2\pi n_k ,
    \qquad n_k \in \mathbb{Z},
    \label{eq:pbr_phase}
\end{equation}
where $d$ is the one-way distance, $c$ is the speed of light, and $n_k$ is the unknown number of elapsed wave cycles. By computing the phase difference $\Delta\theta$ between measurements taken at frequencies separated by $\Delta f$, the propagation distance can be estimated as
\begin{equation}
    d = \frac{c\Delta\theta}{4\pi\Delta f},
    \qquad
    d < d_{\max} = \frac{c}{2\Delta f}.
    \label{eq:pbr_distance}
\end{equation}
Because a single tone pair is highly susceptible to multipath fading and noise, \ac{CS} hop across the full set of $K$ channels, spanning roughly $80$~MHz, to jointly estimate the phase slope:
\begin{equation}
    \hat{d} = \frac{c}{4\pi}
    \arg\min_{s}\sum_{k=1}^{K}\bigl(\theta_k - s f_k\bigr)^2 .
    \label{eq:pbr_slope}
\end{equation}
Returning the \ac{PCT} data to the initiator over the \ac{LE} data connection significantly bottlenecks the overall measurement rate \cite{11207662}, which the inline \ac{PCT} transfer of the Bluetooth Core Specification 6.3 mitigates \cite{ble6.3}.

\subsection{Broadcast based Channel Sounding} \label{sec:broadcast_cs}
Despite the speed improvements of inline \ac{PCT}, maintaining numerous concurrent tag-to-anchor connections continues to suffer from high overhead, energy use, and collision rates. To bypass these limitations, a non-standard, connectionless broadcasting approach to localization was proposed in \cite{10.1145/3842657.3842787}. Much like traditional \ac{RSSI} broadcasting, the tag transmits tones that are independently received by passive anchors. The term broadcast is used loosely here. The tag runs the \ac{CS} tone transmission already available in the \ac{BLE} stack with the reflector replies disabled, so the approach is non-standard in configuration but adds no new on-air signal. Because the available \ac{BLE} bandwidth limits the timing resolution obtainable from direct arrival-time estimation, accurate spatial information is instead extracted from the phase progression across frequencies using \ac{PDoA}.

In standard two-way ranging, the bidirectional exchange cancels the unknown local oscillator phases. As a passive listener takes no part in this exchange, the tag's initial phase cannot be eliminated in the same manner. Instead, it must be canceled spatially by calculating the phase difference across a pair of synchronized anchors observing the identical transmission. For a transmitted tone at frequency $f$ with an initial phase $\phi_0^f$, the phase received at anchor $i$ is given by
\begin{equation}
    \hat\phi_i^f = \phi_0^f - \frac{2\pi f d_i}{c}
    + \phi_{\mathrm{LO},i}^f + 2\pi f \tau_i \pmod{2\pi},
    \label{eq:anchor_phase}
\end{equation}
where $d_i$ is the tag-to-anchor distance, $\phi_{\mathrm{LO},i}^f$ is the phase offset introduced by the anchor's \ac{LO} at tone $f$, and $\tau_i$ is its residual timing error relative to the backbone reference, with the anchor keeping local time $t - \tau_i$. The spatial phase difference $\hat\phi_{i,j}^f \triangleq \hat\phi_i^f - \hat\phi_j^f$ between a synchronized receiver pair cancels the initial tag phase $\phi_0^f$, as both anchors capture the exact same transmission:
\begin{equation}
    \hat\phi_{i,j}^f = -\frac{2\pi f}{c} d_{i,j}
    + \Delta\phi_{\mathrm{LO}}^f + 2\pi f \tau_{i,j} \pmod{2\pi},
    \label{eq:anchor_combined}
\end{equation}
where $d_{i,j} \triangleq d_i - d_j$. As in standard \ac{PBR}, stepping the frequency in increments of $\Delta f$ resolves the modulo-$2\pi$ ambiguity, yielding the spatial range difference:
\begin{equation}\label{eq:dist}
\begin{split}
    \hat d_{i,j} &= -\frac{c}{2\pi \Delta f}\Delta\hat\phi_{i,j} \\
    &= d_{i,j} - c\tau_{i,j} - \frac{c}{2\pi\Delta f}
    \bigl(\Delta\phi_{\mathrm{LO}}^{f+\Delta f}-\Delta\phi_{\mathrm{LO}}^{f}\bigr).
\end{split}
\end{equation}
If each anchor derives its tone \acp{LO} coherently from a common backbone reference, the \ac{LO} difference term becomes frequency-independent and vanishes. This leaves a range-difference estimate biased solely by the residual synchronization error, of magnitude $c|\tau_{i,j}|$. Geometrically, each synchronized pair defines a hyperbola passing through the tag's position, with $N$ anchors providing $N-1$ independent spatial constraints.

\subsection{Limitations of Broadcasting} \label{sec:limitations}

The phase model in \eqref{eq:anchor_phase} assumes a single \ac{LOS} propagation path between the tag and the anchor. However, in indoor environments, the tone arriving at anchor $i$ is the superposition of the direct path and a set of reflections. Expressing the received sample as a complex observation, and not only as its extracted phase, the signal becomes
\begin{equation}
\begin{split}
    z_i(f) = H_i(f)
    e^{j\left(\phi_0^f + \phi_{\mathrm{LO},i}^f + 2\pi f \tau_i\right)},  \\
    H_i(f) = \sum_{\ell=0}^{L_i-1} \alpha_{i,\ell}
    e^{-j 2\pi f \tau_{i,\ell}},
\end{split}
    \label{eq:multipath_channel}
\end{equation}
where $\alpha_{i,\ell}$ and $\tau_{i,\ell}$ denote the complex gain and delay of the $\ell$-th path, with $\tau_{i,0} = d_i/c$ representing the direct path. Consequently, the measured quantity $\hat\phi_i^f = \arg z_i(f)$ is the phase of a superposition, and no single distance $d_i$ satisfying \eqref{eq:anchor_phase} exists. Multipath invalidates the phase model entirely.

This effect is dictated by the available bandwidth. Two paths with a delay difference smaller than $1/B$ are inseparable for any estimator operating on a $B$-wide observation. Across the $K$ \ac{CS} channels, $B \approx 80$~MHz, resulting in a physical path-length resolution of $c/B \approx 3.8$~m. First-order reflections in standard rooms have excess path lengths well below this threshold, and are blended into the direct path. For a direct path and a single reflection with relative gain $\rho_i = \alpha_{i,1}/\alpha_{i,0}$ and excess delay $\Delta\tau_i$, the channel response simplifies to
\begin{equation}
    H_i(f) = \alpha_{i,0} e^{-j2\pi f \tau_{i,0}}
    \left(1 + \rho_i e^{-j2\pi f \Delta\tau_i}\right).
    \label{eq:two_path}
\end{equation}
Crucially, the excess term $\arg\left(1+\rho_i e^{-j2\pi f\Delta\tau_i}\right)$ is not linear in $f$. The phase slope estimated by \eqref{eq:pbr_slope} therefore deviates from the true direct-path delay $\tau_{i,0}$, biasing the resulting range difference. Specular reflections are a deterministic function of room geometry and tag position. The bias is correlated across anchors and remains constant while the tag is stationary, and it persists across repeated measurements. This constitutes an accuracy floor, largely dictating how multipath must be modeled, as addressed in Section~\ref{sec:evaluation}.

Beyond multipath, the broadcasting architecture introduces a strict synchronization cost. Standard two-way \ac{CS} is self-referenced, as the same device originates and receives the exchange, so its clock offset cancels identically. Conversely, while the spatial difference in \eqref{eq:dist} cancels the tag phase $\phi_0^f$, it leaves the anchors' local references intact, resulting in the residual term $c|\tau_{i,j}|$. This term is additive in the range domain and physically indistinguishable from a genuine path-length difference. A mere $1$~ns inter-anchor timing residual contributes $30$~cm of range-difference error, irrespective of the estimator used. Furthermore, the deployment must ensure that the \ac{LO} difference vanishes. If each anchor re-acquires its \ac{LO} phase per frequency hop, $\Delta\phi_{\mathrm{LO}}^f$ is no longer frequency-independent and enters \eqref{eq:dist} scaled by $c/(2\pi\Delta f) \approx 48$~m per radian of error at $\Delta f = 1$~MHz. It is therefore useful to separate $\tau_{i,j}$ into a static component, which can be calibrated at commissioning, and a per-tone random component. The magnitudes of these requirements are quantified in Section~\ref{sec:requirements}.

\section{Proposed Architecture} \label{sec:architecture}
The accuracy floor established in Section~\ref{sec:limitations} originates in the estimator and not in the underlying measurement. Unlike standard phase-slope fitting, subspace methods decompose a frequency response into its propagation paths. These methods are constrained by the \ac{SNR} and model order rather than the physical bandwidth, making the earliest resolved arrival the direct path. Applying such a method would recover $\tau_{i,0}$ directly, bypassing the blended slope of \eqref{eq:two_path}. However, a subspace decomposition requires a pure channel response of the form $H_i(f)$ as defined in \eqref{eq:multipath_channel}, which a passive anchor never observes. Its only usable observable is the pairwise spatial ratio, canceling the unknown tag phase:
\begin{equation}
    \frac{z_i(f)}{z_j(f)}
    = \frac{\sum_{\ell} \alpha_{i,\ell} e^{-j2\pi f \tau_{i,\ell}}}
        {\sum_{\ell} \alpha_{j,\ell} e^{-j2\pi f \tau_{j,\ell}}}
    e^{j\left(\Delta\phi_{\mathrm{LO}}^f + 2\pi f \tau_{i,j}\right)}.
    \label{eq:ratio_obstruction}
\end{equation}
This ratio of exponential sums is not itself a sum of exponentials, and allows no subspace decomposition. A passive anchor suffers more than a suboptimal estimator. It lacks a recoverable channel for estimation. The proposed \ac{RA-PDoA} architecture extends the broadcast scheme specifically to provide this missing channel.

\subsection{Active and Passive Anchors} \label{sec:roles}

\begin{figure*}
    \centering
    \includegraphics[width=0.975\textwidth]{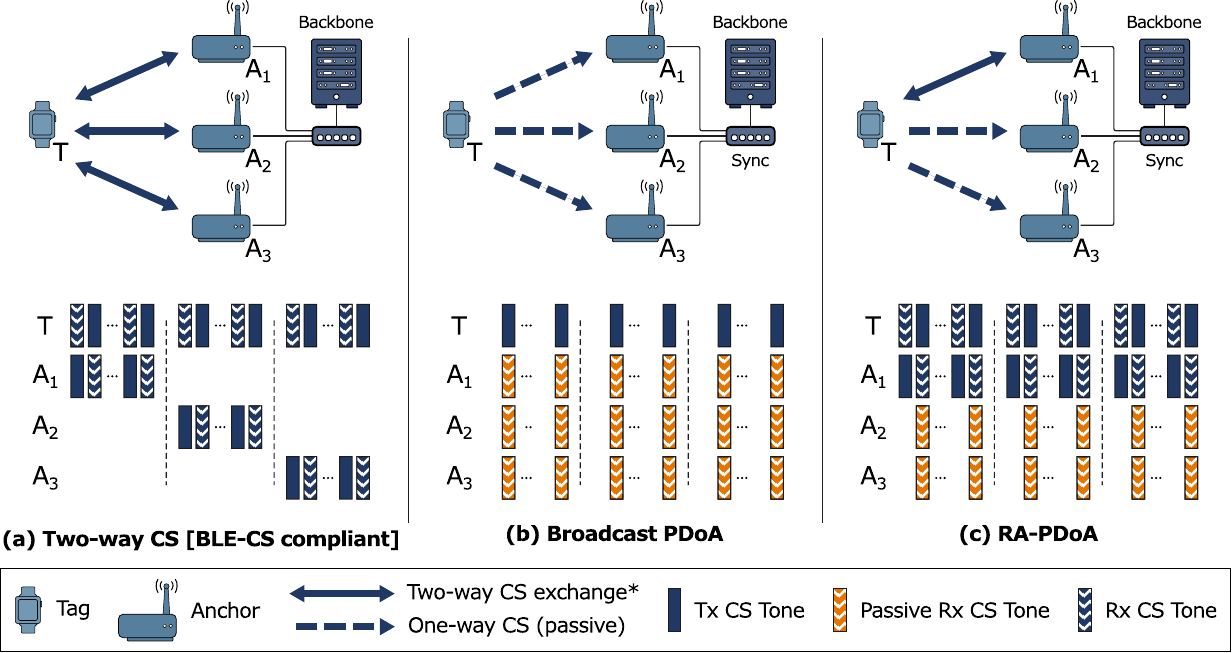}
    \caption{Localization topologies using Bluetooth Channel Sounding. (a) Fully standard-compliant two-way Channel Sounding, where each anchor performs a ranging procedure sequentially. (b) Broadcast \ac{PDoA}, connectionless broadcast to synchronized passive anchors. (c) \ac{RA-PDoA}, combining one active two-way \ac{CS} link with passive reception at the other anchors. $(*)$ The two-way exchange follows the standardized \ac{CS} procedure, whereas passive capture and transmit-only operation do not.}
    \Description[The three different localization topologies using Bluetooth Channel Sounding]{The biggest difference is the direction of the data flow and the time scheduling. With the RA-PDoA being a hybrid version.}
    \label{fig:architecture}
\end{figure*}

Each localization cycle comprises two components, illustrated alongside the two reference architectures in Figure~\ref{fig:architecture}. One anchor, denoted the active anchor $A$, maintains a standard two-way \ac{CS} exchange with the tag, obtaining the round-trip response $z_a^{(2)}(f)$. Additionally, the $N-1$ anchors capture the same tones and yield the one-way observations $z_i(f)$ of \eqref{eq:multipath_channel}. 
The passive anchors do not transmit or participate in separate ranging exchanges. Instead, they receive the tones exchanged between the tag and active anchor using a known frequency-hop schedule and transmission timing. The active anchor configures the \ac{CS} procedure with the tag, and therefore knows both. It distributes them over the synchronized backbone, making them available to the passive anchors. This receive-only operation keeps the airtime per localization cycle independent of $N$. 

This results in a hybrid of the two approaches described in Section~\ref{sec:system_model}. Standard \ac{CS} localization requires $N$ sequential two-way exchanges per fix and is fully compliant but scales poorly. Broadcast \ac{PDoA} requires a single transmission and scales, but abandons the two-way exchange entirely and with it the absolute range measurement. \ac{RA-PDoA} keeps the single transmission of the broadcast scheme by letting the passive anchors observe one two-way exchange. The arrangement thus yields three observables per cycle: the round-trip response $z_a^{(2)}(f)$ at the active anchor, which is self-referenced and free of the tag phase $\phi_0^f$, the one-way responses $z_i(f)$ at all $N-1$ anchors, each carrying $\phi_0^f$, and a coarse arrival timestamp at each anchor, referenced to the backbone. Section~\ref{sec:estimation} describes how a position is obtained from them.

\subsection{Channel Recovery at Passive Anchors} \label{sec:recovery}

The limitation identified in \eqref{eq:ratio_obstruction} is that the passive observable is a ratio of channel responses rather than a channel response. The active anchor resolves this obstruction, as its two-way exchange observes a genuine channel directly. Being self-referenced, the round-trip measurement contains neither the tag phase nor the anchor's own oscillator and timing offsets,
\begin{equation}
    z_a^{(2)}(f) = H_a^2(f) + n_a^{(2)}(f),
    \label{eq:twoway_obs}
\end{equation}
so that $H_a^2(f)$ is known to within noise. Writing the anchor-side residue of \eqref{eq:anchor_combined} as $\Delta_{i,a}(f) \triangleq \Delta\phi_{\mathrm{LO}}^f + 2\pi f\tau_{i,a}$, the pairwise ratio taken against the active anchor is
\begin{equation}
    \frac{z_i(f)}{z_a(f)}
    = \frac{H_i(f)}{H_a(f)} e^{j\Delta_{i,a}(f)} .
    \label{eq:ratio_active}
\end{equation}

The unknown $H_i(f)$ can be recovered as $H_i = (z_i/z_a)H_a$ with $H_a = \sqrt{z_a^{(2)}}$. A complex quantity has two square roots differing in sign, so recovering $H_a$ from $z_a^{(2)}$ requires selecting a branch at every tone and maintaining continuity across the comb. This selection is unreliable at frequencies where the response approaches the origin, and a single incorrect choice propagates to all subsequent tones. We therefore square the relation, removing the ambiguity entirely,

\begin{align}
\begin{split}
    \left(\frac{z_i(f)}{z_a(f)}\right)^{2} z_a^{(2)}(f)
    &= \frac{H_i^2(f)}{H_a^2(f)} e^{j2\Delta_{i,a}(f)} \, H_a^2(f) \\
    &= H_i^2(f) e^{j2\Delta_{i,a}(f)} ,
\end{split}
    \label{eq:recovery}
\end{align}
where each quantity on the left is measured. The recovered response $\tilde H_i^2(f) \triangleq H_i^2(f)e^{j2\Delta_{i,a}(f)}$ is a sum of complex exponentials in $f$ and allows the subspace decomposition that \eqref{eq:ratio_obstruction} denied. It is a round-trip response, which is the form a two-way ranging estimator expects. The corresponding conversion from delay to distance is given in Section~\ref{sec:estimation}.

Two properties follow. First, the squaring leaves the synchronization residue unamplified. The term $2\Delta_{i,a}(f)$ contributes a delay offset of $2\tau_{i,a}$ in the round-trip domain, which the subsequent halving returns to a range bias of $-c\tau_{i,a}$, identical to the one already present in the broadcast estimate of \eqref{eq:dist}. The architecture inherits the synchronization requirement without worsening it. Second, $z_a(f)$ appears in the denominator of \eqref{eq:ratio_active}, so noise on the active link is amplified into each recovered response. A weak or obstructed active anchor degrades all $N-1$ passive estimates simultaneously, which is why Section~\ref{sec:selection} treats the assignment of the active role.

\subsection{Active Anchor Selection} \label{sec:selection}

The active anchor has three key roles. It provides the phase reference for each pair difference, it delivers the only absolute range measurement in the system, and through \eqref{eq:ratio_active} its observation appears in the denominator of each recovered response. Its selection is therefore essential. An obstruction affecting a passive anchor $i$ corrupts one of the $N-1$ constraints, while an obstruction affecting the active anchor corrupts all of them and additionally amplifies noise into each recovered channel.
The information required for this selection is already available. Each anchor receives the tag's transmissions, so its received power from the previous cycle is known. The active role is therefore assigned to the anchor observing the tag most strongly,
\begin{equation}
    a = \arg\max_{i} \frac{1}{K}\sum_{k=1}^{K} \left| z_i(f_k) \right|^2 .
    \label{eq:selection}
\end{equation}
The sensitivity of the selection process is evaluated in Section~\ref{sec:ablation}.

\subsection{Practical Implementation Considerations} \label{sec:implementation}
RA-PDoA requires only limited modifications beyond standard BLE \ac{CS} operation. Unlike broadcast \ac{PDoA}, which requires transmit-only CS operation, the tag and active anchor execute an unmodified two-way \ac{CS} procedure with reflector replies enabled. The only non-standard operation is passive reception at the remaining anchors. Because the standard \ac{CS} interface does not expose samples from a procedure in which a device does not participate, these anchors require receiver- or firmware-level access to the captured tones. Importantly, this requires neither a new on-air waveform nor an additional radio. The channel-hopping order of a \ac{CS} procedure is derived from security parameters shared by initiator and reflector \cite{ble6}. The active anchor must therefore distribute the procedure configuration over the backbone, extending the trust boundary to the anchor infrastructure. The evaluation assumes this access, a synchronized backbone, and a known transmit epoch, and therefore assesses feasibility in simulation rather than a complete hardware implementation.

\section{Position Estimation} \label{sec:estimation} 

\begin{figure*}[t]
    \centering
    \begin{tikzpicture}[
        font=\footnotesize,
        >={Stealth[length=1.8mm]},
        blk/.style={draw, rounded corners=1pt, align=center,
                    minimum height=8mm, inner xsep=2pt, text width=18mm},
        inp/.style={blk, fill=black!6},
        dec/.style={blk, fill=black!3},
        sum/.style={draw, circle, inner sep=0.5pt, minimum size=4.5mm},
        ph/.style={->, semithick},
        tm/.style={->, semithick, dashed},
        lb/.style={font=\scriptsize, inner sep=1.5pt},
        eqn/.style={font=\scriptsize}
    ]
    \draw[fill=black!7] (-1.11,2) rectangle (1.1,-0.6);

    \node[inp] (act) at (0,1.4)  {Active anchor two-way};
    \node[inp] (pas) at (0,0.02) {Passive anchor $i$ one-way};

    \node[blk] (rec) at (2.87,0.6)  {Channel recovery};
    \node[eqn, above=0cm of rec] {\eqref{eq:recovery}};

    \node[blk] (slp) at (5.82,1.68)  {Slope fit};
    \node[eqn, above=0cm of slp] {\eqref{eq:dist}};

    \node[blk] (mus) at (5.82,-0.42) {MUSIC};
    \node[eqn, above=0cm of mus] {\eqref{eq:sr_range}};

    \node[blk] (ord) at (3.37,-1.46) {MDL};
    \node[eqn, above=0cm of ord] {\eqref{eq:mdl}};

    \node[blk] (tst) at (2.6,-3.1) {Arrival\\timestamp};
    \node[eqn, above=0cm of tst] {\eqref{eq:toa_model}};

    \node[sum] (sum) at (7.7,1.7) {$+$};
    \node[eqn, below=0cm of sum] {\eqref{eq:slope_absolute}};

    \node[dec] (sw)  at (9.6,0.6)  {Per-link selection};
    \node[eqn, above=0cm of sw] {\eqref{eq:switch}};
    \node[lb, below=0pt of sw] {$\eta$};

    \node[dec] (vet) at (12.1,0.6) {Consistency\\veto};
    \node[eqn, above=0cm of vet] {\eqref{eq:veto}};
    \node[lb, below=0pt of vet] (node1) {$\gamma$};

    \node[blk] (bnd) at (14.6,-3.1) {Bounding region };
    \node[eqn, below=0.5pt of bnd] {\eqref{eq:toa_box}};

    \node[blk] (sol) at (14.6,0.6) {Bounded least squares };
    \node[eqn, above=0pt of sol] {\eqref{eq:solve}};

    \draw[ph] (pas.east) -- ++(0.5,0) |- (1.9,0.46)
        node[lb, above, xshift=5pt, at start] {$z_i$};
    \draw[ph] (act.east) -- ++(0.5,0) |- (1.9,0.81)
        node[lb, above, xshift=5.5pt, pos=0.22] {$z_a$};

    \draw[ph] (act.north) -- ++(0,0.97) -- (7.7,2.8) -- (sum.north)
        node[lb, above, pos=0.47] {};

    \draw[ph] (3.84,0.8) -- ++(0.5,0) |- (slp.west);
    \draw[ph] (3.84,0.4) -- ++(0.5,0) |- (mus.west);
    \draw[ph, arrows=-Stealth] (ord.east) -- ++(1.47,0) -- (mus.south);

    \draw[ph] (slp) -- (sum) node[lb, midway, above] {$\hat d_{i,a}$};
    \draw[ph] (sum.east) -- ++(0.29,0) |- (8.63,0.7)
        node[lb, above, pos=0.18, xshift=11.5pt] {$\hat d_i^{\mathrm{slope}}$};
    \draw[ph] (mus.east) -- ++(1.43,0) |- (8.63,0.37)
        node[lb, below, pos=0.33, xshift=7pt] {$\hat d_i^{\mathrm{sr}}$};

    \draw[ph] (sw) -- (vet) node[lb, midway, above] {$\hat d_i$};
    \draw[ph] (vet) -- (sol);

    \draw[tm] (pas.south) |- (tst.west);
    \draw[tm, semithick] (tst.east) -| (11.6,0.17)
        node[lb, pos=0.25, above] {$\hat d_i^{\mathrm{toa}}$};
    \draw[tm] (tst.east) -- (bnd.west);
    \draw[tm] (bnd) -- (sol);

    \draw[ph] (sol.east) -- ++(0.7,0) node[lb, right] {$\hat{\mathbf p}$};

      \node[draw=none, node font=\tiny] at (0.3,-2.06) {ToA};
  \node[draw=none, node font=\scriptsize] at (14.8,-1.29) {$\mathcal{B}$};
    \end{tikzpicture}
    \caption{\ac{RA-PDoA} position estimation pipeline per anchor. Dashed arrows mark the timing component, while the governing equation is displayed for each component.}
    \Description[Our positioning estimation pipeline per anchor]{Each anchor goes through the same pipeline, where multiple components are used to achieve one final range.}
    \label{fig:pipeline}
\end{figure*}
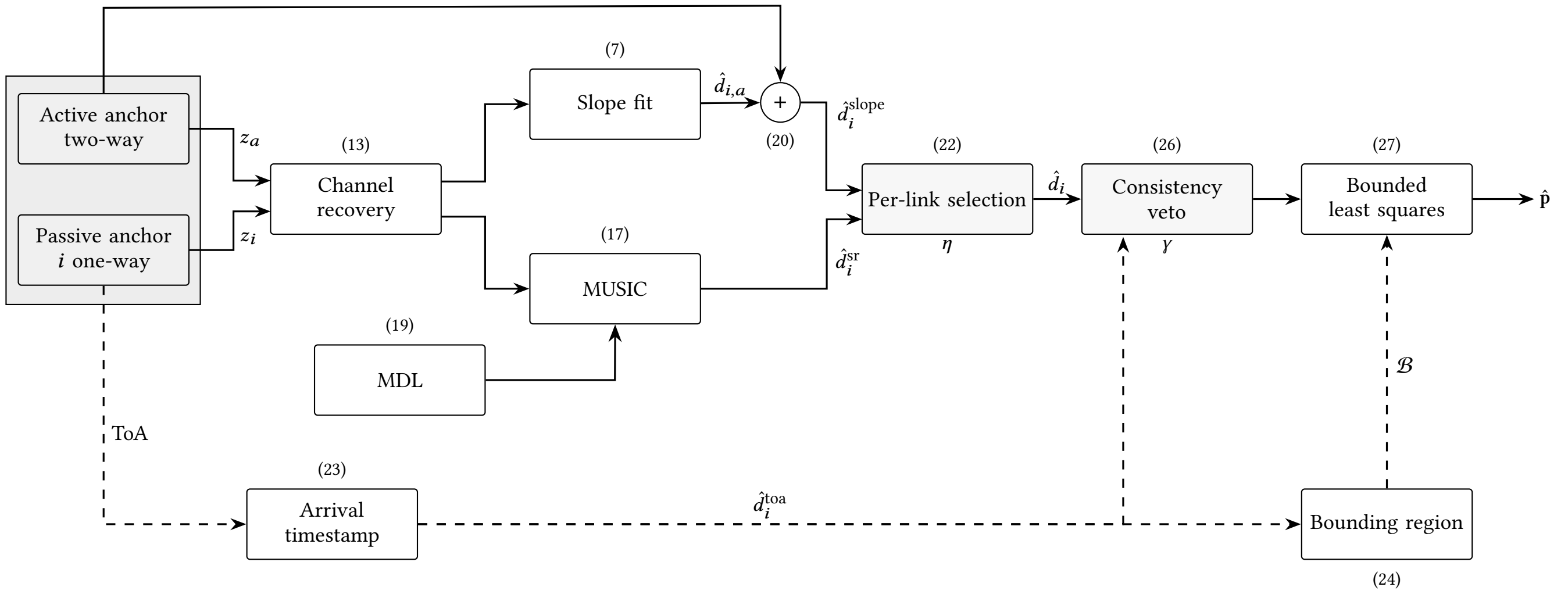

\ac{RA-PDoA} yields three observables per measurement cycle: the recovered round-trip response $\tilde H_i^2(f)$ at each anchor, the pairwise phase differences of \eqref{eq:anchor_combined}, and a coarse arrival timestamp. This section describes how a position estimate is formed from them, and the resulting pipeline is summarized in Figure~\ref{fig:pipeline}. Two range estimates are available per anchor, with complementary failure modes, and the coarse timestamp serves both to constrain the solution and to arbitrate between them.

\subsection{Delay Estimation} \label{sec:delay_estimation}
    
The recovered response is sampled at the $K$ tone frequencies, giving $\tilde{\mathbf h}_i = [\tilde H_i^2(f_1),\dots,\tilde H_i^2(f_K)]^{T}$. The propagation delays are estimated from this vector using \ac{MUSIC}. Subspace decomposition conventionally requires a covariance matrix averaged over repeated snapshots, whereas a passive anchor observes each transmission exactly once. Decomposition is enabled by mapping $\tilde{\mathbf h}_i$ into a Hankel matrix $\mathbf G_i \in \mathbb{C}^{M \times Q}$ of sliding windows along the frequency axis, with pencil parameter $M = \lfloor K/2 \rfloor$ and $Q = K - M + 1$ sub-apertures. Smoothing is performed along frequency and not across snapshots, making the estimator applicable to a single transmission.
Due to the mutual coherence of the direct path and its reflections, forward-backward spatial smoothing is additionally applied to restore the rank. Following \cite{17496}, the augmented covariance is
\begin{equation}
    \mathbf R_i = \frac{1}{2Q}\left(
        \mathbf G_i \mathbf G_i^{H}
        + \mathbf J \left(\mathbf G_i \mathbf G_i^{H}\right)^{*} \mathbf J
    \right),
    \label{eq:covariance}
\end{equation}
where $\mathbf J$ denotes the exchange matrix and $(\cdot)^{*}$ complex conjugation. An eigenvalue decomposition $\mathbf R_i = \mathbf V_i \boldsymbol\Lambda_i \mathbf V_i^{H}$ partitions the eigenvectors into signal and noise subspaces  $[\mathbf V_{\mathrm{s},i}, \mathbf V_{\mathrm{n},i}]$, the dimension of the former being the estimated model order $\hat L_i$ obtained as described in Section~\ref{sec:model_order}.
For a candidate delay $\tau$, the steering vector is $\mathbf e(\tau) = [1, e^{-j2\pi\tau\Delta f},\dots, e^{-j2\pi\tau (M-1)\Delta f}]^{T}$. While \ac{MUSIC} conventionally tests the orthogonality of $\mathbf e(\tau)$ against the noise subspace, the projection identity of \cite{10092848} permits the equivalent evaluation against the smaller signal subspace,
\begin{equation}
    P_i(\tau) = \frac{1}{M - \left\| \mathbf V_{\mathrm{s},i}^{H}
    \mathbf e(\tau) \right\|^{2}} .
    \label{eq:pseudospectrum}
\end{equation}
The resolution of $P_i(\tau)$ is governed by the \ac{SNR} and the accuracy of $\hat L_i$ rather than by the bandwidth $B$, thereby removing the limitation identified in Section~\ref{sec:limitations}.

A reflected path is longer than the direct path, so the direct-path delay corresponds to the earliest maximum of $P_i(\tau)$ rather than the largest. Under the blending of \eqref{eq:two_path} the largest may correspond to a reflection. Let $\mathcal{P}_i$ denote the set of maxima with a prominence above a fixed threshold and a delay above a lower bound that excludes the zero-delay artifact. The delay is then estimated as
\begin{equation}
    \hat\tau_i = \min \mathcal{P}_i ,
    \qquad
    \hat d_i^{\mathrm{sr}} = \tfrac{1}{2} c \hat\tau_i ,
    \label{eq:sr_range}
\end{equation}
where the factor of one half follows from $\tilde H_i^2$ being a round-trip
response, as established in \eqref{eq:recovery}.

\subsection{Model Order Selection} \label{sec:model_order}

The partition of $\mathbf V_i$ into signal and noise subspaces requires the number of resolvable components $\hat L_i$, and \eqref{eq:pseudospectrum} depends on it directly. A conventional and simple criterion selects the index of the largest consecutive drop in the eigenvalue spectrum,
\begin{equation}
    \hat L_i^{\mathrm{ratio}}
    = \arg\max_{1 \le k < M}  \frac{\lambda_k}{\lambda_{k+1}} ,
    \label{eq:ratio_order}
\end{equation}
with $\lambda_1 \ge \dots \ge \lambda_M$ the eigenvalues of $\mathbf R_i$. This presumes a clear separation between signal and noise eigenvalues. Coherent indoor multipath lacks such a separation. Reflections of graded strength yield an eigenvalue spectrum that decays smoothly, and the largest ratio then frequently occurs at a low index, causing the criterion to under-estimate the order.

The consequence extends beyond a loss of resolution. When $\hat L_i < L_i$, the retained signal subspace does not span the true signal space, and the unmodeled path directions remain within the assumed noise subspace $\mathbf V_{\mathrm{n},i}$. As \eqref{eq:pseudospectrum} exhibits a maximum only where the steering vector is orthogonal to that subspace, a path lying inside it produces no maximum at all. An under-estimated order can remove the direct path from the pseudospectrum entirely. In such a scenario the earliest detected maximum belongs to a reflection, and \eqref{eq:sr_range} returns a range that is too long by the corresponding excess path length.

Let $\bar\lambda_{\mathrm{g}}(k)$ and $\bar\lambda_{\mathrm{a}}(k)$ denote the geometric and arithmetic means of the $M-k$ smallest eigenvalues $\{\lambda_{k+1},\dots,\lambda_M\}$. Their ratio is unity precisely when those eigenvalues are equal, the case where the corresponding subspace may be attributed to white noise, and decreases as they become disparate. The \ac{MDL} criterion \cite{1164557} penalizes departure from this case against the cost of each additional parameter,
\begin{equation}
    \mathrm{MDL}(k) = -\tilde Q(M-k)
    \log \frac{\bar\lambda_{\mathrm{g}}(k)}{\bar\lambda_{\mathrm{a}}(k)}
    + \tfrac{1}{2}k(2M-k)\log \tilde Q ,
    \label{eq:mdl}
\end{equation}
with $\hat L_i = \arg\min_k \mathrm{MDL}(k)$ and $\tilde Q = 2Q$, the sub-aperture count doubled by the forward-backward averaging in \eqref{eq:covariance}.
On synthetic channels with known ranges, \ac{MDL} reduces the median range error of the eigenvalue ratio from $9.1$ to $4.2$~cm and the $90$th percentile from $30.9$ to $16.9$~cm (Appendix~\ref{app:model_order}). Section~\ref{sec:ablation} quantifies the cost of the \ac{MDL} criterion in position error.

\subsection{Per-Link Estimator Selection} \label{sec:switch}

Two range estimates are available at each passive anchor. The subspace estimator of \eqref{eq:sr_range} yields $\hat d_i^{\mathrm{sr}}$ directly. The slope fit of Section~\ref{sec:system_model} yields a range difference $\hat d_{i,a}$, which the absolute range of the active anchor makes absolute,
\begin{equation}
    \hat d_i^{\mathrm{slope}} = \hat d_a + \hat d_{i,a}.
    \label{eq:slope_absolute}
\end{equation}
The two-way link serves both estimators, not only the channel recovery of \eqref{eq:recovery}. Neither estimate dominates the other. Where the single-path model of \eqref{eq:anchor_phase} holds, the slope fit is generally preferable, as it uses all $K$ tones in a single regression and is subject to neither model-order error nor peak selection, two sources of variance that the subspace estimator is unable to avoid. Where the model fails, the slope estimate is biased in the manner established in Section~\ref{sec:limitations}, and the subspace estimator is required. A single fix may combine anchors with clear \ac{LOS} and anchors with heavily reflected paths. Consequently, the choice must be made per link.

A decision statistic is used, under the single-path model the pair-difference phase is linear in frequency, and departure from linearity indicates that the slope fit should not be trusted. Denoting by $\hat\beta_i$ and $\hat\psi_i$ the least-squares slope and intercept fitted to the unwrapped pair-difference phase, the residual
\begin{equation}
    \varepsilon_i = \left[
    \frac{1}{K}\sum_{k=1}^{K}
    \left( \arg\left(\frac{z_i(f_k)}{z_a(f_k)}\right)
    - \hat\beta_i f_k - \hat\psi_i \right)^{2}
    \right]^{1/2}
    \label{eq:residual}
\end{equation}
measures how poorly the single-path model describes that link. The estimator is then selected per anchor as
\begin{equation}
    \hat d_i =
    \begin{cases}
        \hat d_i^{\mathrm{sr}}, & \varepsilon_i > \eta ,\\[2pt]
        \hat d_i^{\mathrm{slope}}, & \text{otherwise} ,
    \end{cases}
    \label{eq:switch}
\end{equation}
requiring no knowledge of the reflectivity of the environment and no additional computation. The active anchor does not rely on \eqref{eq:switch}. It observes its own channel directly, with neither reconstruction nor division by the response of another anchor, allowing the subspace estimate to be used unconditionally. Selection of the threshold $\eta$ is addressed in Section~\ref{sec:result}.

\subsection{Coarse Time of Arrival} \label{sec:toa}

Each anchor records the \ac{ToA} of the tag's transmission relative to the backbone, yielding a third observable. The active anchor distributes the scheduled transmit epoch over the backbone. We assume that each anchor's static clock offset has been calibrated at commissioning. Let $\tau_i$ denote the remaining per-cycle synchronization error after calibration, bounded by $|\tau_i| \leq \bar\tau_i$. After subtracting the calibrated offset, the corrected \ac{ToA} range is
\begin{equation}
    \hat d_i^{\mathrm{toa}} = d_i + \delta_i - c\tau_i + \xi + \nu_i ,
    \qquad \nu_i \sim \mathcal{N}\left(0, (c\sigma_t)^2\right),
    \label{eq:toa_model}
\end{equation}
where $\delta_i \geq 0$ is the excess path of an obstructed direct path, $\xi$ is
the transmit-epoch error discussed below, and $\sigma_t$ is the timestamping
uncertainty. The sign of $\tau_i$ follows from the clock
convention of \eqref{eq:anchor_phase}, where anchor $i$ keeps local time
$t - \tau_i$. This clock adds the $+2\pi f\tau_i$ phase term, an apparent range
of $d_i - c\tau_i$, and a timestamp on the same clock reads $\tau_i$ early, so
the \ac{ToA} range is shortened by the same $c\tau_i$. The tag transmits relative to
the packet of the active anchor, so its actual transmit time differs from the
scheduled epoch by the propagation delay $d_a/c$ and the turnaround time of the
tag. The first term is estimated from the two-way range of the active anchor and
the second is set by the configured \ac{CS} step timing, but neither is exact.
The remaining error $\xi$, expressed in range units, is common to all anchors
and is not negligible. In the evaluated configuration, the active-range error
alone has a median of $17$~cm and a $90$th percentile of $41$~cm, comparable to
$c\sigma_t = 30$~cm. The evaluation assumes a known transmit epoch, with
$\xi = 0$, so its $\sigma_t$ covers the timestamp noise only and the reported
\ac{ToA} results are optimistic with respect to this term. The \ac{ToA} precision is meter-class
and far inferior to either phase-based estimate, making it inadequate for
positioning in its own right. Timestamp noise is independent of phase-measurement
noise, and multipath affects the \ac{ToA} differently. However, the residual timing
error $\tau_i$ and the excess path $\delta_i$ also affect phase-derived ranges, so the
\ac{ToA} and phase estimates are not fully independent. We account for this shared timing error in the
feasible region and consistency threshold. The \ac{ToA} therefore serves as a coarse,
differently affected check, analogous to the role of \ac{RTT} in standard
\ac{CS}. The first use is to constrain. Each measurement confines the tag to a disc around its anchor, and the intersection of these discs
\begin{equation}
    \mathcal{B} = \bigcap_{i=1}^{N}
    \left\{ \mathbf p : d_i(\mathbf p)
    \leq \hat d_i^{\mathrm{toa}} + \kappa c\sigma_t + \bar\xi + c\bar\tau_i \right\}
    \label{eq:toa_box}
\end{equation}
bounds the admissible positions, with $d_i(\mathbf p)$ the three-dimensional
distance of \eqref{eq:geometry}, $\kappa$ setting the timestamp-noise
confidence level, $\bar\xi$ bounding the transmit-epoch error, and $c\bar\tau_i$
covering the bounded residual timing error.
Only the outer bound is applied, as the excess path $\delta_i$ lengthens the
\ac{ToA} by an unknown amount and would invalidate an inner bound.
The axis-aligned bounding box of $\mathcal{B}$ is provided to the solver of
Section~\ref{sec:solution}. This constraint is unable to improve accuracy, as it
is far looser than the phase-based estimates it bounds. However, it limits
divergence when the range estimates are inconsistent, provided the scheduled
transmit epoch and residual timing bounds are valid.

The second use is to arbitrate. The selection rule of \eqref{eq:switch} decides
whether the single-path model adequately describes a link, but it cannot detect
whether the subspace estimator selected a reflected path. Under the phase model,
the subspace range has the form
\begin{equation}
    \hat d_i^{\mathrm{sr}} = d_i + \delta_i - c(\tau_i - \tau_a) + e_i^{\mathrm{sr}},
    \label{eq:sr_timing_error}
\end{equation}
where $a$ is the active anchor and $e_i^{\mathrm{sr}}$ is the estimator error.
The excess path $\delta_i$ and the residual $\tau_i$ enter both estimates identically
and cancel in their difference. The discrepancy between the \ac{ToA} and phase
estimates therefore contains only the timestamp noise, the transmit-epoch error
$\xi$, the residual timing error of the active anchor, and the estimator error.
The veto is
\begin{equation}
    \left| \hat d_i^{\mathrm{sr}} - \hat d_i^{\mathrm{toa}} \right| > \gamma
    \Longrightarrow
    \hat d_i \leftarrow \hat d_i^{\mathrm{slope}} .
    \label{eq:veto}
\end{equation}

The threshold $\gamma$ must exceed the largest legitimate discrepancy between
the corrected \ac{ToA} and phase estimates, including residual timing error, and
remain below the excess path length of a false peak. For example, at confidence
level $\kappa$, a sufficient lower bound is
$\gamma > \kappa c\sigma_t + \bar\xi + c\bar\tau_a + e_{i,\max}^{\mathrm{sr}}$,
where $\bar\xi$ bounds the transmit-epoch error, $\bar\tau_a$ bounds the active
anchor's residual timing error, and $e_{i,\max}^{\mathrm{sr}}$ bounds legitimate
subspace-estimator error. The evaluation sets $\bar\xi = 0$, consistent with its
known transmit epoch.

\subsection{Position Solution} \label{sec:solution}
Sections~\ref{sec:switch} and~\ref{sec:toa} yield an absolute range estimate $\hat d_i$ at each anchor, including the passive ones. This constitutes a qualitative change from the broadcast baseline, where only the $N-1$ range differences of \eqref{eq:dist} are observable. The position is now obtained by circular trilateration rather than hyperbolic multilateration, avoiding the poor conditioning associated with hyperbolic geometry.
However, the recovered ranges are not independent in their errors. An unresolved reflection blends with the direct path and displaces the estimated delay later, as established in \eqref{eq:two_path}. Due to the shared reflecting surfaces, each anchor is displaced in the same direction, and a component common to all ranges translates the solution and survives the least-squares fit. Such a common term is structurally equivalent to a receiver clock bias in satellite navigation, and the same treatment applies. It is estimated jointly with the position, preventing any corruption of the result. Using the modeled distance $d_i(\mathbf p)$ of
\eqref{eq:geometry},
\begin{equation}
    \left(\hat{\mathbf p},\, \hat b\right)
    = \arg\min_{\mathbf p \in \mathcal{B}, b}
    \sum_{i=1}^{N} \left( d_i(\mathbf p) + b - \hat d_i \right)^{2} ,
    \label{eq:solve}
\end{equation}
with $\mathcal{B}$ the bounding region of \eqref{eq:toa_box}. The common offset $b$ also absorbs the component of the synchronization residual of Section~\ref{sec:limitations} that is shared, and is discarded once the position is obtained.

The additional unknown carries a cost. With $N$ absolute ranges and three unknowns the problem retains $N-3$ degrees of redundancy, which equals the redundancy available to the $N-1$ differences and two unknowns of the baseline. The benefit of \eqref{eq:solve} derives from representing the coherent error explicitly, and from the circular geometry that the absolute ranges allow, and not from additional measurements. The problem is solved numerically subject to the bounds, with the initial estimate taken as the center of $\mathcal{B}$.

\section{Evaluation Methodology} \label{sec:evaluation}

\subsection{Simulator} \label{sec:simulator}

The evaluation uses a purpose-built Python-based Monte Carlo simulator implementing the \ac{CS} physical-layer model of Section~\ref{sec:system_model}. The objective is not to reproduce a specific commercial implementation, but to isolate the physical and architectural mechanisms that determine the feasibility of \ac{RA-PDoA}. Rather than representing ranging error as an aggregate noise process, the simulator synthesizes the complex per-tone observation $z_i(f)$ at each anchor from the modeled propagation, hardware, and backbone impairments. Position estimates are then obtained using the same processing pipeline that would be applied in a deployment. The modeled mechanisms and their default settings are listed in Table~\ref{tab:params}.

Generative AI tools from Anthropic were used during the development of the simulator and the analysis of its results. Claude Fable~5 assisted in writing parts of the simulator model. Claude Opus~4.8 and Claude Sonnet~4.6 assisted in writing the evaluation scripts and were used for initial data processing and figure generation, providing a high-level overview of the simulation results. This exploratory processing was subsequently cleaned up by the authors, and the reported results and figures were produced from the cleaned-up scripts. The same two models were also used as an additional check of the equation derivations. All generated code, processed data, and derivations were reviewed and verified by the authors, who take full responsibility for the content of this paper.

\begin{table}
    \caption{Modeled mechanisms and default configuration}
    \label{tab:params}
    \centering
    \setlength{\tabcolsep}{2pt}
    \begin{tabular}{ll}
        \toprule
        \multicolumn{2}{l}{\textbf{Deployment}} \\
        \midrule
        Room / anchors      & $10\times10$~m, $4$ corner-mounted \\
        Anchor / tag height & $2.8$~m / $1.2$~m \\
        Tag positions       & $16$ ($4\times4$ grid) \\
        Trials per position & $500$ \\
        \midrule
        \multicolumn{2}{l}{\textbf{Signal}} \\
        \midrule
        Tone sweep           & $79$ channels, $\Delta f = 1$~MHz, $B = 78$~MHz \\
        Tone dwell          & $50~\mu$s \\
        Link budget         & free space, $-20$~dBm, NF $6$~dB \\
        SNR ceiling    & $30$~dB \\
        \midrule
        \multicolumn{2}{l}{\textbf{Channel}} \\
        \midrule
        Multipath           & first-order image source, $|\Gamma|$ swept \\
        Body shadowing      & $60^\circ$ half-angle, $15$--$30$~dB, \\
                            & $0.2$--$1.0$~m excess \\
        \midrule
        \multicolumn{2}{l}{\textbf{Backbone}} \\
        \midrule
        Sync offset         & $0$ (swept, Section~\ref{sec:requirements}) \\
        Per-tone jitter     & $0$ (swept, Section~\ref{sec:requirements}) \\
        Oscillator phase    & coherent \\
        Hardware group delay & $0$ (swept, Section~\ref{sec:requirements}) \\
        Anchor survey error & $0$ (swept, Section~\ref{sec:requirements}) \\
        \midrule
        \multicolumn{2}{l}{\textbf{Estimator}} \\
        \midrule
        Timestamp uncertainty & $\sigma_t = 1$~ns \\
        \ac{ToA} box confidence & $\kappa = 3$ \\
        Residual timing bound & $\bar\tau_i = \kappa \times$ sync offset std \\
        Switch / veto threshold & $\eta = 0.15$~rad, $\gamma = 2.0$~m \\
        \bottomrule
        \end{tabular}
\end{table}

\subsection{Multipath Model} \label{sec:multipath_model}

\begin{figure}
    \centering
    \includegraphics[width=\columnwidth]{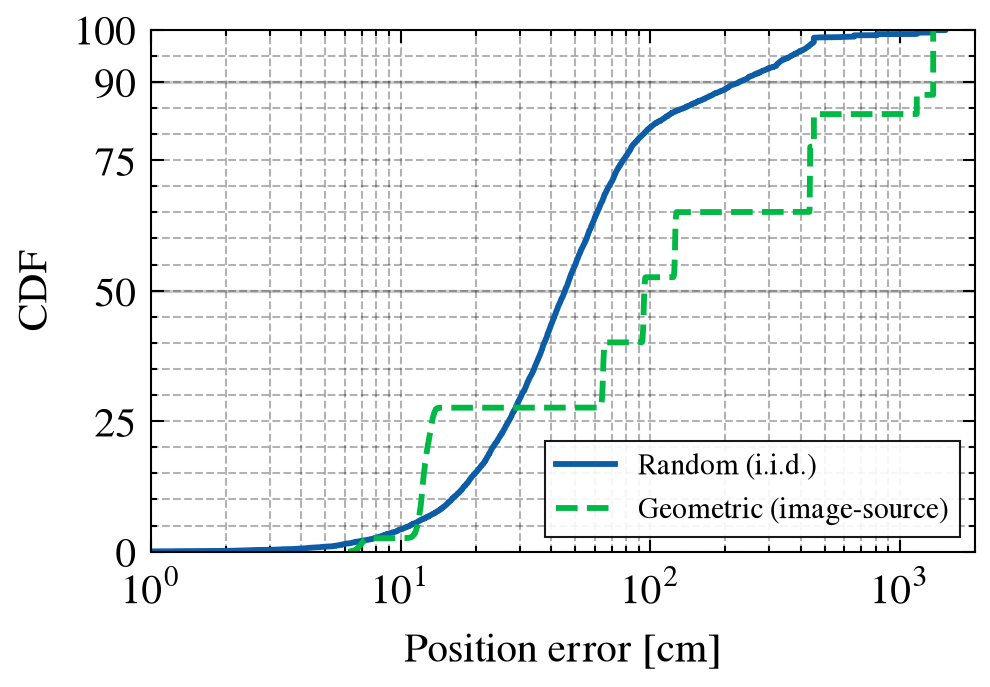}
    \caption{Error distribution under statistical and image-source multipath models.}
    \label{fig:multipath}
    \Description[Random is smooth CDF, geometric is not]{First 25 percent the geometric has the lowest error impact however from there on it has a higher error impact compared to random multipath.}
\end{figure}

Multipath is generated by first-order image-source ray tracing against the room's walls. The alternative, drawing reflections independently per anchor and per repetition, is common in the literature and is also implemented, but the two are not interchangeable. Independently drawn reflections are uncorrelated across anchors and redrawn at each repetition, so their contribution averages away both in the least-squares fit and over repeated measurements. Specular reflections from real surfaces behave differently. They are a deterministic function of room geometry and tag position, identical at each repetition while the tag is stationary, and correlated across anchors due to the same surfaces illuminating all of them.

Figure~\ref{fig:multipath} shows that the resulting difference is substantial. Under identical conditions, the statistical model yields a median error of $45.6$~cm and a $90$th percentile of $228$~cm, while image-source reflections in the same room yield $94.7$~cm and $1360$~cm, with the fraction of diverged estimates rising from $6.2\%$ to $34.9\%$. A statistical multipath model understates the median by a factor of two and the tail by a factor of six, concealing the failures that occur in practice. For this reason we adopt the image-source model throughout.

\subsection{Simulation Design}

The default deployment is given in Table~\ref{tab:params}: a single room with ceiling-mounted corner anchors and a chest-worn tag, evaluated at $16$ positions on a regular grid with $500$ independent trials at each. Individual experiments vary specific parameters from this default. Wall reflectivity is swept over $|\Gamma| \in \{0, 0.2, 0.4, 0.6\}$ throughout, spanning an anechoic room to one bounded by materials such as glass and concrete. Body shadowing is evaluated both enabled and disabled, as it and multipath are separate mechanisms with different remedies. The active-anchor study additionally varies anchor count and layout, using a circular arrangement so that the count may be changed without simultaneously changing the geometry. Section~\ref{sec:requirements} reintroduces each backbone impairment individually. Unless stated otherwise, all are zero, ensuring that the multipath and shadowing terms are isolated.

\subsection{Metrics and Baseline}

Accuracy is reported as the median and the \ac{CEP90} of the position error over all tag positions and trials. The median characterizes typical performance and the $90$th percentile the tail, which for a positioning system is frequently the binding requirement. Additionally, the fraction of estimates reaching the solver bounds is reported, distinguishing degraded accuracy from outright divergence.
The baseline throughout is the broadcast \ac{PDoA} scheme of Section~\ref{sec:broadcast_cs}, evaluated with a fixed reference anchor. Adaptive reference selection was also assessed and did not improve it. The fixed configuration is its strongest, and is the one reported in Section~\ref{sec:result}.
Standard two-way \ac{CS} localization is also computed, both with the slope estimator and with the subspace estimator of Section~\ref{sec:delay_estimation}, separating the contributions of the architecture and of the estimator. Two constraints follow from the sequential nature of its exchanges. It uses trilateration, and it receives no coarse \ac{ToA}.
The switch threshold $\eta$ of \eqref{eq:switch} is the only parameter selected from the data. It is calibrated on a single seed and reported on three further seeds, ensuring that the quoted performance is that of a threshold chosen without access to the data used for its evaluation.

\subsection{Assumptions}

First, the tag's per-tone phase $\phi_0^f$ is set to zero in the Monte Carlo runs. Since it cancels exactly in \eqref{eq:anchor_combined} irrespective of its value, this affects no result, and the cancellation is verified separately. Second, the active-anchor criterion of \eqref{eq:selection} is evaluated from the true received power and not a noisy estimate, which is optimistic to the extent that a practical implementation would average \ac{RSSI} over a finite number of packets. Third, reflections are restricted to first-order specular paths from the vertical surfaces of the room. Diffuse scattering, furniture, and higher-order paths are not modeled. Fourth, the transmit epoch is assumed known, with $\xi = 0$ in Section~\ref{sec:toa}, and $\sigma_t$ covers the timestamp noise only. The active-range error that would drive $\xi$ is comparable to $c\sigma_t$, so the \ac{ToA} results are optimistic in this respect.

\section{Results and Discussion} \label{sec:result} 

\subsection{Accuracy across Reflectivity and Shadowing} \label{sec:accuracy}

\begin{figure}
    \centering
    \includegraphics[width=\columnwidth]{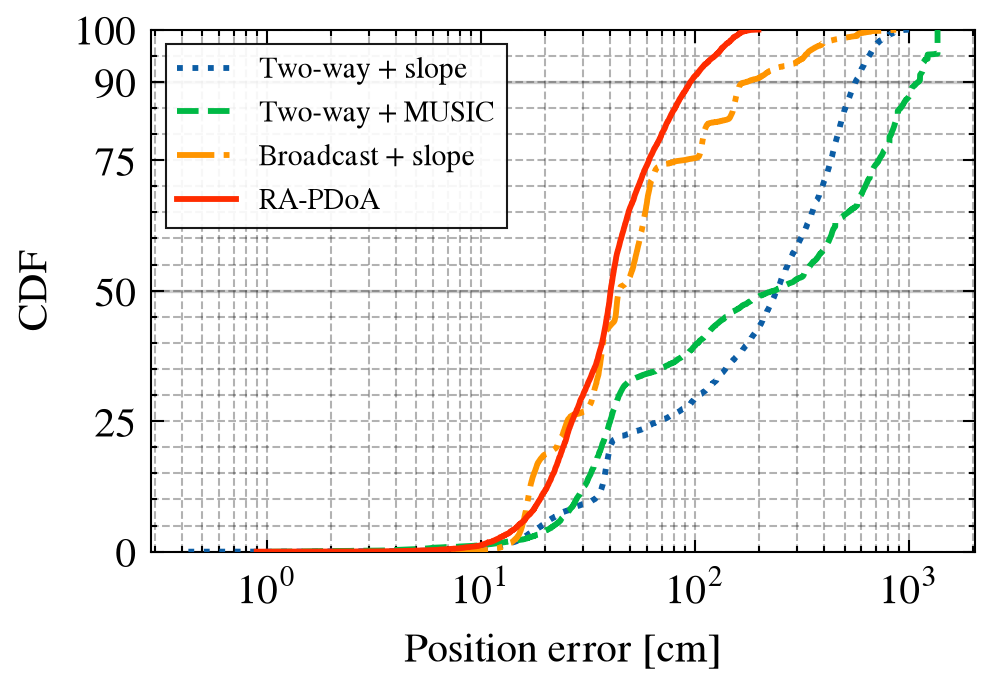}
    \caption{Error distribution at the default operating point ($|\Gamma| = 0.4$, body shadowing enabled).}
    \label{fig:cdf}
    \Description[RA-PDoA is the best]{Significant difference between broadcasting based and sequential two-way, with RA-PDoA being the best.}
\end{figure}

\begin{table}
    \caption{Position error between \ac{BLE} Two-way \ac{CS} (slope), \ac{BLE} Two-way \ac{CS} (MUSIC), Broadcast \ac{PDoA} (slope) and proposed \ac{RA-PDoA} expressed in cm, given as median (90th percentile)}
    \label{tab:accuracy}
    \centering
    \compacttable
    \setlength{\tabcolsep}{4pt}
    \begin{tabular}{lrrrr}
        \toprule
        $|\Gamma|$
        & \makecell{\ac{BLE} Two-way\\\acs{CS} (slope)}
        & \makecell{\ac{BLE} Two-way\\\acs{CS} (\acs{MUSIC})}
        & \makecell{Broadcast\\\acs{PDoA}}
        & \makecell{\textbf{\acs{RA-PDoA}}} \\
        \midrule
        \multicolumn{5}{l}{\textit{Body shadowing disabled}} \\
        0.0 & 0.3 (0.6)     & 0.6 (1.3)      & 0.5 (1.0)     & \textbf{0.5 (1.0)}   \\
        0.2 & 16.7 (21.6)   & 17.4 (50.3)    & 18.2 (26.0)   & \textbf{16.5 (24.1)} \\
        0.4 & 35.5 (40.3)   & 35.9 (84.7)    & 36.6 (76.9)   & \textbf{32.4 (69.2)} \\
        0.6 & 94.8 (285.8)  & 12.0 (40.3)    & 299.7 (401.3) & \textbf{17.3 (64.1)} \\
        \midrule
        \multicolumn{5}{l}{\textit{Body shadowing enabled}} \\
        0.0 & 295.3 (645.4) & 352.3 (1169.1) & 36.0 (76.4)   & \textbf{34.4 (58.1)} \\
        0.2 & 267.0 (620.3) & 328.0 (1169.1) & 35.3 (90.5)   & \textbf{31.5 (69.5)} \\
        0.4 & 241.6 (564.1) & 234.0 (1106.7) & 44.1 (175.3)  & \textbf{40.4 (95.5)} \\
        0.6 & 274.9 (634.5) & 71.8 (980.1)   & 167.7 (413.6) & \textbf{35.5 (93.6)} \\
        \bottomrule
    \end{tabular}
\end{table}

Table~\ref{tab:accuracy} reports the median and \ac{CEP90} horizontal error for all four architectures across the swept reflectivity, with body shadowing enabled and disabled.
Super-resolution provides robustness to multipath in both architectures. Without body shadowing at $|\Gamma| = 0.6$, replacing the slope fit with the subspace estimator improves the two-way median from $94.8$ to $12.0$~cm and the broadcast median from $299.7$ to $17.3$~cm. However, it is not uniformly preferable. Applied unconditionally in the two-way arm, it degrades CEP90 from $21.6$ to $50.3$~cm at $|\Gamma| = 0.2$ and from $40.3$ to $84.7$~cm at $|\Gamma| = 0.4$. Where the single-path model still holds, the slope fit is the better estimator, and model-order and peak-selection error contribute only variance. Detecting this situation is the purpose of the per-link selection of \eqref{eq:switch}. \ac{RA-PDoA} consequently outperforms the two-way subspace arm at $|\Gamma| = 0.4$, despite reconstructing its channels instead of measuring them.

\begin{table}
    \caption{Cost of removing each component, $\Delta$CEP90 in cm}
    \label{tab:ablation}
    \centering
    \compacttable
    \begin{tabular}{lrrrr}
        \toprule
        & \multicolumn{3}{c}{\textbf{Shadowed}} & \textbf{Clear} \\
        \cmidrule(lr){2-4}\cmidrule(lr){5-5}
        \textit{Removed} & $0.2$ & $0.4$ & $0.6$ & $0.4$ \\
        \midrule
        \ac{MDL}   & $-8.0$ & $+26.2$ & $-5.4$ & $+35.2$ \\
        Per-link switch          & $+3.4$ & $+0.1$ & $0.0$ & $+0.5$ \\
        \ac{ToA} veto            & $+1.2$ & $+29.3$ & $+82.8$ & $+11.2$ \\
        \ac{ToA} veto and bound  & $+1.4$ & $+36.3$ & $+170.2$ & $+14.9$ \\
        Common-bias unknown      & $+6.8$ & $-6.1$ & $+25.5$ & $-21.5$ \\
        Active-anchor selection  & $+7.0$ & $+6.0$ & $+4.3$ & $0.0$ \\
        \midrule
        Full \ac{RA-PDoA}  & $69.5$ & $95.5$ & $93.6$ & $69.2$ \\
        \bottomrule
    \end{tabular}
\end{table}

The fraction of estimates reaching the solver bounds also changes what the errors in Table~\ref{tab:accuracy} mean. The two-way subspace arm reaches the solver bounds on $32$ to $47\%$ of shadowed estimates, so its reported CEP90 describes the extent of the search region and not a measured error, while an unbounded solver would return arbitrarily large values in such cases. The underlying failure occurs on individual links and is gross. A body-blocked anchor attenuated by $30$ to $60$~dB over the round trip presents the subspace estimator with noise, yielding an essentially arbitrary delay, and a single such range suffices to drive a trilateration to its bounds. The slope-fit arm saturates far less while performing uniformly worse, reflecting the same distinction. It is consistently biased, while the subspace arm is occasionally catastrophic.

This observation has a practical consequence beyond the comparison. A two-way deployment is unable to rely on its ranges alone, as it has no means of recognizing that one of them is meaningless. Each anchor produces a range, and nothing in the measurement distinguishes a good one from a link measuring noise. Thus, some form of outlier rejection is a precondition for practical use, and it must draw on information beyond the ranging itself. \ac{RA-PDoA} obtains this structurally, through two mechanisms. The coarse \ac{ToA}, though far less precise than either phase-based estimate, has independent timestamp noise and a differently affected multipath error, and \eqref{eq:veto} uses it to reject a range that no other measurement supports. Additionally, the assignment of \eqref{eq:selection} places the single two-way link on the least obstructed anchor, largely avoiding the scenario that produces such ranges.

\ac{RA-PDoA} is the only configuration that remains usable in all evaluated scenarios, and the only one with estimates that never reach the solver bounds. The error distribution at that operating point is shown in Figure~\ref{fig:cdf}. At the default operating point of $|\Gamma| = 0.4$ with body shadowing its CEP90 is $95.5$~cm $[93.4, 98.0]$, compared to $175.3$~cm $[162.4, 196.6]$ for the broadcast baseline, a factor of $1.84$, with saturation falling from $3.6\%$ to $0.0\%$. Standard two-way \ac{CS} remains the more accurate scheme where links are unobstructed and reflectivity is moderate, at $40.3$ compared to $69.2$~cm of CEP90 at $|\Gamma| = 0.4$, as it measures each channel directly rather than recovering $N-1$ of them by division. That accuracy is purchased with $N$ sequential exchanges per fix and is lost entirely once the wearer's body obstructs an anchor.
The worst case for the proposed estimator occurs at intermediate reflectivity, as the common-bias unknown of \eqref{eq:solve} absorbs only the error shared across anchors. Moderate reflections bias each anchor differently, while strong reflections make all anchors read long by a similar amount.

The single tuned parameter, the switch threshold $\eta$ of \eqref{eq:switch}, was calibrated on one random seed and evaluated on three further seeds not used in its selection. Across those held-out seeds the worst case over all reflectivities is $1.00\times$ the baseline with body shadowing and $1.05\times$ without, indicating that \ac{RA-PDoA} does not degrade accuracy relative to the baseline in any evaluated scenario.

\subsection{Estimator Component Contributions} \label{sec:ablation}

Table~\ref{tab:ablation} reports the cost of removing each mechanism from the complete configuration. Leave-one-out is preferred over a cumulative build-up due to the interaction between the mechanisms.
No mechanism is uniformly beneficial, and the scenarios where each contributes are consistent with its purpose. The model-order criterion is worth $26$~cm of CEP90 at the default operating point and $35$~cm without body shadowing, the conditions where an underestimated order removes the direct path from the signal subspace. It is marginally harmful at $|\Gamma| = 0.2$, where few reflections are strong enough for that to occur, and at $|\Gamma| = 0.6$ with body shadowing. The latter observation is consistent with the decomposition of Section~\ref{sec:accuracy}. An underestimated order merges the direct path with its reflection and biases all anchors late together, producing an error that is largely common and absorbed by \eqref{eq:solve}, while a correct order leaves a smaller residual with a greater differential proportion.

The per-link switch is worth $8.5$~cm of median error at $|\Gamma| = 0.2$ and essentially nothing beyond it. This is the expected behavior, as the switch exists to route links to the slope fit where the single-path model still holds, and by $|\Gamma| = 0.4$ almost none do. Its value lies in preventing the degradation that unconditional subspace estimation causes at low reflectivity, visible in the two-way arm of Table~\ref{tab:accuracy}.

The coarse \ac{ToA} contributes in both of its roles, and increasingly with reflectivity. Removing the consistency veto costs $29$~cm of CEP90 at $|\Gamma| = 0.4$ and $83$~cm at $|\Gamma| = 0.6$, while removing the bounding region as well costs a further $87$~cm at $|\Gamma| = 0.6$. The two act at different points. The veto prevents an individual erroneous range from entering the solution, while the bound prevents the solution from diverging when more than one does.

The common-bias unknown provides the clearest example of a mechanism that carries a cost. It is worth $31$~cm of median error at $|\Gamma| = 0.6$ with body shadowing, where the range error is largely common across anchors, but costs $21$~cm of CEP90 at $|\Gamma| = 0.4$ without shadowing, where it is not. The additional unknown consumes a degree of freedom unsupported by the data, and the common share of the range error is lowest in precisely that scenario. Enabling it conditionally on an estimate of that share may be a natural refinement, and is not attempted here.

\begin{table}
    \caption{Active-anchor assignment, with body shadowing,
    median and CEP90 in cm}
    \label{tab:selection}
    \centering
    \compacttable
    \begin{tabular}{lrrrr}
        \toprule
        & \multicolumn{2}{c}{\textbf{$|\Gamma| = 0.2$}} & \multicolumn{2}{c}{\textbf{$|\Gamma| = 0.4$}} \\
        \cmidrule(lr){2-3}\cmidrule(lr){4-5}
        \textit{Assignment} & $\tilde x$ & $90\%$ & $\tilde x$ & $90\%$ \\
        \midrule
        Fixed, anchor $0$   & $30.5$ & $76.5$ & $41.5$ & $101.5$ \\
        Fixed, anchor $1$   & $42.2$ & $92.6$ & $40.2$ & $110.4$ \\
        Fixed, anchor $2$   & $39.8$ & $80.5$ & $39.8$ & $101.7$ \\
        Fixed, anchor $3$   & $35.6$ & $60.9$ & $39.7$ & $93.0$ \\
        \midrule
        Strongest broadcast & $31.5$ & $69.5$ & $40.4$ & $95.5$ \\
        \bottomrule
    \end{tabular}
\end{table}

The final mechanism is the assignment of the active anchor, reported separately in Table~\ref{tab:selection} as it is a property of the deployment and not of the estimator. The choice is less important than the three roles of the active anchor would suggest. Across the four fixed assignments the median error varies by a factor of $1.38$ at $|\Gamma| = 0.2$ and by $1.09$ or less elsewhere. The mechanisms above account for this behavior, with the consistency veto, the bounding region, and the common-bias unknown each limiting the damage a poor assignment can do. The received-power rule of \eqref{eq:selection} falls short of the best fixed assignment, trailing it by about $1$~cm of median error, but it avoids the worst without requiring knowledge of which anchor is favorable. Such knowledge is unavailable at deployment time, as the best assignment depends on the room and on the distribution of tag positions, and differs between them. An oracle selecting on true ranging error does not bound the achievable performance and is worse than the rule in two scenarios, as the anchor with the most accurate range is not necessarily the one that yields the best position.

Applied to the baseline the same criterion is harmful, raising the CEP90 of the broadcast scheme from $90.5$ to $148.2$~cm at $|\Gamma| = 0.2$ and its median from $44.1$ to $55.6$~cm at $|\Gamma| = 0.4$. Received power identifies an obstructed link but carries no information about multipath bias, and in a purely differential scheme the error of the reference anchor propagates to all $N-1$ constraints. The criterion is useful in \ac{RA-PDoA} due to the subspace estimator resolving the multipath that power alone leaves invisible, and offers no benefit where the estimator is unable to do so.

\subsection{Anchor Count and Placement} \label{sec:anchors}

\begin{table}
    \caption{Anchor count and placement, with body shadowing enabled. Median
    ($\tilde x$) and CEP90 ($90\%$) horizontal error in cm against wall
    reflectivity. GDOP is the mean geometric dilution of precision over the
    evaluation grid.}
    \label{tab:anchors}
    \centering
    \compacttable
    \begin{tabular}{lrrrrrrr}
        \toprule
        & & \multicolumn{2}{c}{$|\Gamma| = 0.0$} & \multicolumn{2}{c}{$|\Gamma| = 0.4$}
        & \multicolumn{2}{c}{$|\Gamma| = 0.6$} \\
        \cmidrule(lr){3-4}\cmidrule(lr){5-6}\cmidrule(lr){7-8}
        Layout & GDOP & $\tilde x$ & $90\%$ & $\tilde x$ & $90\%$ & $\tilde x$ & $90\%$ \\
        \midrule
        $4$, circle & $1.40$ & $52.8$ & $89.2$ & $40.0$ & $133.8$ & $43.8$ & $150.4$ \\
        $6$, circle & $1.19$ & $46.2$ & $98.4$ & $49.2$ & $138.0$ & $55.6$ & $137.6$ \\
        $8$, circle & $1.01$ & $39.0$ & $73.0$ & $45.4$ & $135.0$ & $50.1$ & $132.7$ \\
        \midrule
        $4$, corner & $0.91$ & $34.4$ & $58.1$ & $40.4$ & $95.5$ & $35.5$ & $93.6$ \\
        \bottomrule
    \end{tabular}
\end{table}

Geometric considerations predict that accuracy should improve with anchor count, as the mean geometric dilution of precision falls from $1.40$ at four circularly arranged anchors to $1.01$ at eight. 
Table~\ref{tab:anchors} reports the measured behavior in the realistic channel. The predicted trend is weak and not monotonic. On the circular layout the median at $|\Gamma| = 0.4$ moves from $40.0$ to $49.2$ to $45.4$~cm as the count increases from four to eight. The four-anchor corner deployment outperforms all of them, at $40.4$~cm median and $95.5$~cm CEP90 compared to $45.4$ and $135.0$~cm for eight circular anchors. Appendix~\ref{app:anchors} shows that i.i.d.\ multipath restores most of the predicted trend and that the finding holds on a denser $5 \times 5$ grid, indicating that the failure to scale is a property of the multipath model of Section~\ref{sec:multipath_model} and not of anchor count.

Placement substitutes for count. Four corner-mounted anchors achieve a geometric dilution slightly better than eight arranged on a circle, $0.91$ compared to $1.01$, and outperform them substantially once multipath is present. These findings suggest that a deployment is better served by placing few anchors well than by adding more of them, which inverts the usual expectation that infrastructure density buys accuracy. No natural corner-equivalent exists at six or eight anchors, so placement and count are not fully separated here. The circular family isolates count, while the corner deployment establishes what placement alone can achieve.

\subsection{Alternative Mechanisms} \label{sec:negative}

We evaluated three alternative strategies for mitigating erroneous range estimates: link exclusion, robust position estimation, and phase-based consistency gating. Neither replacement for the coarse-\ac{ToA} veto matches it. With body shadowing, a robust loss reaches a CEP90 of $115.6$~cm at $|\Gamma| = 0.4$, compared to $95.5$~cm with the veto, and slope-based gating reaches $261.7$~cm at $|\Gamma| = 0.6$, compared to $93.6$~cm. Link exclusion on top of the veto lowers the CEP90 at $|\Gamma| = 0.4$ only from $95.5$ to $92.0$~cm while raising the median from $40.4$ to $42.6$~cm. Appendix~\ref{app:negative} reports the detailed results.
Overall, these results show that effective consistency validation requires a measurement with independent noise and different failure modes, rather than a second estimate derived from the same phase observations.

\section{Infrastructure Requirements} \label{sec:requirements}

Section~\ref{sec:limitations} identified a synchronization requirement as the cost of broadcast operation, and the preceding results idealized it. This section reintroduces each backbone error term and quantifies what the deployment must provide. Under the broadcast baseline the multipath floor exceeds each accuracy target before any impairment is added, so the requirements are unreachable and not merely demanding.

\subsection{Error Budget}

Table~\ref{tab:budget} reports the contribution of each impairment, measured as the increase in CEP90 over the corresponding baseline under common random numbers, so that a term with no effect measures as no effect. Confidence intervals are obtained by paired bootstrap, and a contribution with an interval spanning
zero is reported as indistinguishable from none.

\begin{table}
    \caption{Error budget. Baseline CEP90 and the contribution of each impairment, in cm}
    \label{tab:budget}
    \setlength{\tabcolsep}{2pt}
    \centering
    \begin{tabular}{lrrrr}
        \toprule
        & \multicolumn{2}{c}{\textbf{Shadowing}} & \multicolumn{2}{c}{\textbf{No Shadowing}} \\
        \cmidrule(lr){2-3}\cmidrule(lr){4-5}
        & Broadcast & \ac{RA-PDoA} & Broadcast & \ac{RA-PDoA} \\
        \midrule
        Baseline CEP90            & $175.3$ & $95.5$ & $76.9$ & $69.2$ \\
        Saturation                & $3.6\%$ & $0.0\%$ & $0.0\%$ & $0.0\%$ \\
        \midrule
        Sync offset, $300$~ps     & $\sim 0$ & $\sim 0$   & $+0.3$  & $+9.4$ \\
        Per-tone jitter, $30$~ps  & $+165.5$ & $+8.9$     & $+84.7$ & $+18.5$ \\
        \ac{LO} uncal., $0.3$~rad & $\sim 0$ & $-8.9$     & $+7.9$  & $+6.9$ \\
        Cable delay, $200$~ps     & $+25.2$  & $+16.3$    & $+5.0$  & $+8.3$ \\
        Anchor survey, $5$~cm     & $\sim 0$ & $+2.5$     & $+2.1$  & $+1.3$ \\
        \bottomrule
    \end{tabular}
\end{table}

Notably, terms that are indistinguishable from zero under the baseline acquire measurable contributions under \ac{RA-PDoA}. These contributions do not indicate a deterioration. The terms were always present, but were concealed beneath a multipath floor several times larger, becoming visible only once that floor is removed.
Per-tone jitter is the one term with a contribution that is reduced and not revealed. Jitter perturbs the phase of each tone independently and degrades a slope fit across the sweep far more than a subspace decomposition, which draws its estimate from the covariance structure and not from a regression on individual tones.
The negative \ac{LO} entry under body shadowing, with an interval of $[-11.0, -7.0]$~cm, is likely an artifact, as oscillator error is unable to improve a range estimate. It does not originate in the mechanism of Section~\ref{sec:model_order}, and it is not relied upon.

\subsection{Timing and Survey Requirements}

\begin{table}
    \caption{Infrastructure requirements: impairment level at which CEP90 reaches
    the stated accuracy target}
    \label{tab:requirements}
    \centering
    \compacttable
    \begin{tabular}{lrrrr}
        \toprule
        & \multicolumn{2}{c}{\textbf{Broadcast}} & \multicolumn{2}{c}{\textbf{\ac{RA-PDoA}}} \\
        \cmidrule(lr){2-3}\cmidrule(lr){4-5}
        Target & $100$~cm & $200$~cm & $100$~cm & $200$~cm \\
        \midrule
        \multicolumn{5}{l}{\textit{Body shadowing disabled}} \\ 
        Sync offset      & $930$~ps & $2.22$~ns & $663$~ps & $1.99$~ns \\
        Per-tone jitter  & $17$~ps & $38$~ps & $\geq 300$~ps & $\geq 300$~ps \\
        \ac{LO} spread   & $0.72$~rad & $1.28$~rad & $0.85$~rad & $\geq 2.00$~rad \\
        Anchor survey    & $25$~cm & $\geq 50$~cm & $36$~cm & $\geq 50$~cm \\
        \midrule
        \multicolumn{5}{l}{\textit{Body shadowing enabled}} \\ 
        Sync offset      & unmet & $470$~ps & $347$~ps & $1.78$~ns \\
        Per-tone jitter  & unmet & $11$~ps  & $<3$~ps  & $\geq 300$~ps \\
        \ac{LO} spread   & unmet & $0.37$~rad & $<0.10$~rad & $\geq 2.00$~rad \\
        Anchor survey    & unmet & $20$~cm  & $8$~cm  & $\geq 50$~cm \\
        \bottomrule
    \end{tabular}
\end{table}

Table~\ref{tab:requirements} reports the level at which each impairment pushes CEP90 through a given accuracy target. An entry of \textit{unmet} indicates that the baseline with that impairment absent already exceeds the target, so no level of it satisfies the requirement. The $50$~cm target, unmet in all configurations, is omitted.
Under the baseline with body shadowing, no impairment level achieves $100$~cm, as the floor alone exceeds it. Under \ac{RA-PDoA} each entry at that target is finite.

Per-tone jitter is the term that relaxes most and is also the most important one. Without body shadowing the $100$~cm requirement moves from $17$~ps to beyond $300$~ps, the upper end of the swept range, so the true tolerance is likely looser still. This constitutes a change of regime and not merely of degree. Sub-nanosecond inter-anchor stability at the tens-of-picoseconds level is difficult to distribute, while hundreds of picoseconds is routine. Jitter is also the term that resists calibration. A static offset, whether in the clock, the cable, or the surveyed anchor position, is measured once at commissioning and subtracted, so the requirement applies to the residual. A per-tone perturbation allows no such treatment.

The sync offset requirement moves in the opposite direction, tightening from $930$ to $663$~ps without body shadowing. With the multipath floor removed, this previously invisible term consumes a measurable share of the margin, making the requirement more demanding due to its increased relevance and not increased fragility. The veto threshold is held at $\gamma = 2.0$~m in every sweep, which falls below the sufficient bound of Section~\ref{sec:toa} above a sync offset of about $1.2$~ns.
The severe shadowed entries at the $100$~cm target reflect limited headroom and not the terms themselves, as the shadowed baseline of $95.5$~cm leaves $4.5$~cm of margin.

\section{Conclusion} \label{sec:conclusion} 
This paper introduced \ac{RA-PDoA}, a scalable \ac{BLE} \ac{CS} localization architecture combining one active anchor with multiple passive, receive-only anchors. The active anchor performs a standard two-way \ac{CS} exchange and provides the reference required to recover a channel response at each passive anchor. This enables super-resolution delay estimation without additional ranging exchanges, keeping the airtime per localization cycle independent of the number of anchors. Passive reception improves robustness to body shadowing, while super-resolution mitigates the multipath bias of phase-slope estimation. Under body shadowing, \ac{RA-PDoA} reduces the 90th-percentile localization error from 175.3~cm to 95.5~cm and eliminates the solver divergence observed for 3.6\% of the broadcast baseline estimates. Coarse \ac{ToA} measurements further improve reliability by providing a consistency check for the phase-derived ranges with independent timestamp noise. The infrastructure analysis shows that mitigating the multipath-induced accuracy floor makes the synchronization requirements for a specified accuracy target quantifiable and substantially increases the allowable per-tone timing jitter in the evaluated conditions. The results constitute a simulation-based feasibility study using first-order specular reflections, a synchronized anchor backbone, and a known transmit epoch. Within these assumptions, it indicates that a single active ranging link is sufficient to make \ac{BLE} \ac{CS} localization both scalable and robust to multipath.

% Acknowledgments are not allowed in the review copy.
\ifanon\else
\begin{acks}
This research was funded by the imec ICON project UBIWAU, financed by imec and Flanders Innovation \& Entrepreneurship.
\end{acks}
\fi

\bibliographystyle{ACM-Reference-Format}
\bibliography{bib}

\appendix

\section{Model-Order Criteria} \label{app:model_order}

This appendix compares the \ac{MDL} criterion of \eqref{eq:mdl} with the eigenvalue ratio of \eqref{eq:ratio_order} and the \ac{AIC}. The related \ac{AIC} criterion \cite{1164557},
\begin{equation}
    \mathrm{AIC}(k) = -2\tilde Q(M-k)
    \log \frac{\bar\lambda_{\mathrm{g}}(k)}{\bar\lambda_{\mathrm{a}}(k)}
    + 2k(2M-k),
    \label{eq:aic}
\end{equation}
which differs only in applying a penalty that does not grow with $\tilde Q$, was also evaluated and is not used. Under the selection rule of \eqref{eq:sr_range}, over-estimation of the order proves to be the more damaging error. The additional components are fitted to noise and generate spurious maxima at arbitrary delays, including delays shorter than the direct path, which the earliest-peak rule then selects. An under-estimated order biases the range estimate long, while an over-estimated one biases it short and does so unboundedly.

\begin{figure}
    \centering
    \includegraphics[width=\columnwidth]{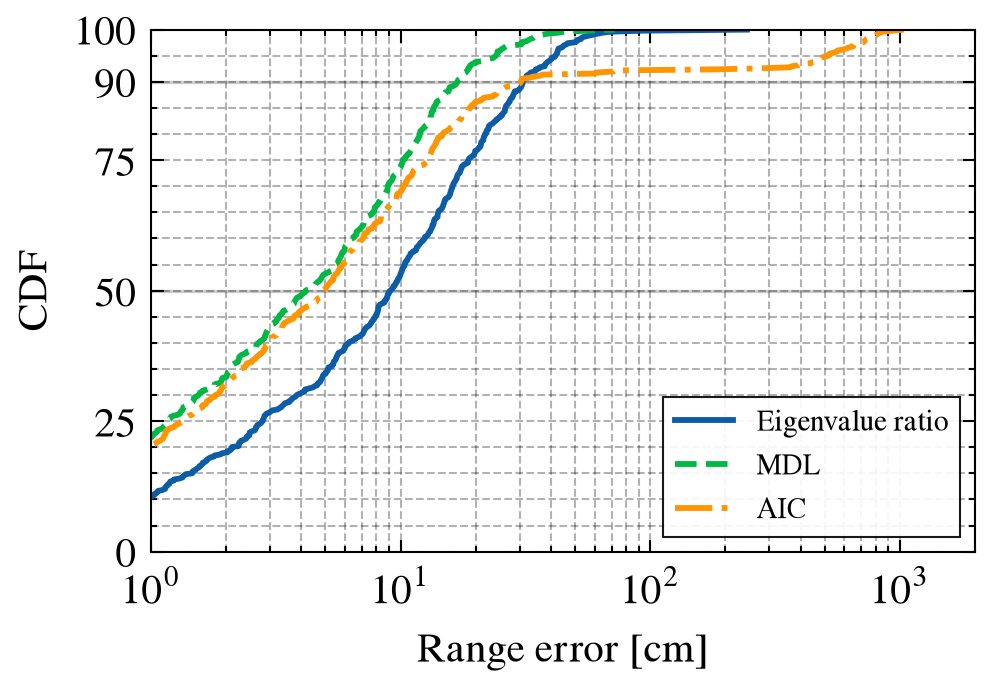}
    \caption{Direct-path range error under the three model-order criteria for $500$ synthetic channels with the true range known by construction.}
    \Description[MDL slightly outperforms the other two]{The range error is more or less consistent across the whole slope with a constant offset between the three methods, except AIC turns worse in the tail.}
    \label{fig:order_error}
\end{figure}

Figure~\ref{fig:order_error} compares the three criteria in isolation. \ac{MDL} halves the range error
relative to the eigenvalue ratio at both the median, from $9.1$ to $4.2$~cm, and the $90$th percentile, from $30.9$ to $16.9$~cm. \ac{AIC} recovers the median improvement, at $4.9$~cm, but not the tail. Its $90$th percentile of $30.3$~cm equals the one of the eigenvalue ratio, and $7.6\%$ of its estimates yield errors above $2$~m, compared to $0.2\%$ and $0.0\%$ for the other two criteria. These results indicate that counting the order correctly is more important than counting it higher. The estimated orders themselves show why. Over the links of the evaluation of Section~\ref{sec:evaluation}, the eigenvalue ratio returns a median order of $2$, compared to $4$ for \eqref{eq:mdl}, and returns a single component on $45\%$ of them, declaring a single-path channel in a room with a geometry that allows four first-order wall reflections.

The improvement in Figure~\ref{fig:order_error} is likely a lower bound on what the criterion contributes within the complete system. The synthetic channels carry no reconstruction noise and draw their excess path lengths uniformly and not from a particular room geometry, so they do not reproduce the meter-scale failures that motivate the criterion. Section~\ref{sec:ablation} quantifies the cost of the \ac{MDL} criterion in position error.

\section{Anchor Count Sensitivity} \label{app:anchors}

The mean geometric dilution of precision over the evaluation grid falls from $1.40$ at four circularly arranged anchors to $1.01$ at eight, and the corresponding Cram\'er--Rao bound improves by a factor of $1.41$. In a channel free of multipath the solve realizes essentially all of it, the measured median improving by $1.40$ over the same range. Neither the geometry nor the estimator is the limitation.

Under a statistical multipath model the effect is largely absent. Repeating the sweep with i.i.d. reflections instead of the image-source model, changing nothing else, restores most of the predicted trend. The median falls from $69.4$ to $56.4$~cm across the same four-to-eight range, an improvement of $1.23$ compared to the $1.41$ that geometry predicts, while the image-source model delivers $0.88$ and thus a degradation. These results indicate that the failure to scale is a property of the multipath model of Section~\ref{sec:multipath_model} and not of anchor count.

The conclusion holds regardless of the tag grid. Repeating both sweeps on a denser $5 \times 5$ grid, sampling the room interior omitted by the evaluation grid, leaves each qualitative finding intact. The statistical model still improves with count, from $66.3$ to $57.7$~cm, and the image-source model still does not, moving from $62.9$ to $62.4$~cm. Furthermore, four corner anchors still outperform eight circular ones, by a wider margin than on the evaluation grid at $36.2$ compared to $62.4$~cm. Only the particular sequence differs, with the denser grid placing its best circular result at six anchors and not at its worst. The individual values move while the absence of a trend does not, which is itself consistent with the underlying mechanism. Each value reflects one specific geometry interacting with one specific set of walls, and the conclusion is carried by the failure to improve and not by a single number.

\section{Alternative Mechanisms} \label{app:negative}

First, unreliable links were excluded based on received power or phase consistency. Received power identifies body-blocked links reliably and reduces the median error from 34.4~cm to 1.5~cm in the absence of multipath. However, at $\lvert\Gamma\rvert=0.4$, excluding blocked links increases the median error from 40.4~cm to 42.6~cm and only marginally improves the 90th-percentile error from 95.5~cm to 92.0~cm. Phase consistency is less discriminative, achieving an area under the receiver operating characteristic of 0.712. With only four anchors, the loss of geometric diversity therefore offsets the benefit of removing an unreliable link.

Second, robust losses were considered in the position solver. Replacing least squares with a Huber or soft-$\ell_1$ loss reduces the 90th-percentile error under body shadowing at $\lvert\Gamma\rvert=0.4$ from 124.8~cm to 115.6~cm, compared with 95.5~cm using coarse-\ac{ToA} validation. Robust losses can downweight large residuals but cannot reliably reject the meter-scale errors caused by incorrect path selection. Explicitly excluding the least-consistent anchor improves the 90th-percentile error to 97.9~cm at $\lvert\Gamma\rvert=0.4$, but degrades performance at higher reflectivity because one quarter of the available anchor geometry is removed.

Third, we evaluated consistency gating between the super-resolution and phase-slope estimates. This approach is ineffective because both estimates originate from the same phase observations and exhibit correlated failures under multipath. At $\lvert\Gamma\rvert=0.6$, slope-based gating increases the 90th-percentile error from 176.4~cm to 261.7~cm. The coarse arrival timestamp is more effective despite its lower precision because it observes the propagation distance with independent timestamp noise and a differently affected multipath error.

\end{document}